\documentstyle[aps,prl,epsf,floats]{revtex}
\def\figuresize{\ifpreprintsty 10cm \else 8cm \fi}

\begin{document}

\ifpreprintsty \else
\twocolumn[\hsize\textwidth\columnwidth\hsize\csname@twocolumnfalse%
\endcsname \fi

\draft
\title{Quantum Coulomb Glass: Anderson localization in an interacting system}

\author{Thomas Vojta}
\address{Materials Science Institute, University
of Oregon, Eugene, OR97403, USA}

\author{Frank Epperlein and Michael Schreiber}
\address{Institut f\"ur Physik, Technische Universit\"at, D-09107 Chemnitz, Germany}
\date{version August 25, compiled \today}
\maketitle

\begin{abstract}

The quantum Coulomb glass model describes disordered interacting electrons
on the insulating side of a metal-insulator transition. By taking quantum 
fluctuations into account it can describe not only the localized limit
but also the weakly localized regime. We discuss several possibilities
to generalize the concept of  Anderson localization to interacting electron
systems such as the quantum Coulomb glass and define criteria for localization.
The corresponding physical quantities are calculated by numerically exact
diagonalization.
The results indicate that 
single-particle excitations
close to the Fermi energy become more strongly localized
under the influence of interaction.

\end{abstract}


\ifpreprintsty \else
] \fi              


\section{Introduction}

Disordered interacting electronic systems such as semiconductor 
impurity bands or granular metals have attracted a large amount of  theoretical and 
experimental interest \cite{reviews}. Even the presence of disorder or interaction alone
leads to interesting phenomena, and the competition between the two gives rise
to a very rich behavior. The single-electron states in non-interacting systems
become localized in space for strong enough disorder. This phenomenon is called
Anderson localization \cite{anderson}; it is responsible for the metal-insulator transition
in non-interacting systems.
The generalization of this concept to interacting systems is not
straightforward since many-particle physical properties cannot in general be 
derived from single-particle properties. Consequently, a
definition of localization in many-particle systems will not be unique.
In this paper we therefore consider several possibilities to generalize the 
concept of Anderson localization to interacting systems and discuss their
relative merits. As an example we then apply the resulting localization criteria
to the quantum Coulomb glass model of disordered insulators and
calculate the corresponding physical quantities by numerically 
exact diagonalization.

The paper is organized as follows: In Sec. II we motivate and define the quantum 
Coulomb glass model. Sec. III is devoted to a discussion of Anderson localization
in interacting systems and the presentation of our results.
Sec. IV contains some discussions and conclusions.

\section{From classical to quantum Coulomb glass}
Almost the entire current understanding of disordered
insulators has been obtained from studying the
insulating limit where the electrons are completely localized 
and can thus be described as
classical point charges (we will denote this regime as the
classical insulating regime from now on). 
Although it is not a quantum mechanical 
system, the classical disordered insulator is a complicated many-body problem.
Pollak predicted \cite{pollak70} an interaction-induced reduction of the 
single-particle DOS at the Fermi
energy in disordered insulators. Later Efros and Shklovskii defined \cite{es75}
the generic model of the classical insulating regime, the
classical Coulomb glass model which consists of
point charges in a random potential which interact via Coulomb interactions.
The model is defined on a regular hypercubic lattice with 
$N=L^d$ ($d$ is the spatial dimensionality) 
sites occupied by $K N$ (spinless) 
electrons ($0\!<\!K\!<\!1$). To ensure charge neutrality
each lattice site carries a compensating positive charge of  $Ke$. The Hamiltonian
of the classical Coulomb glass reads
\begin{equation}
H_{\rm cl} = \sum_i (\varphi_i - \mu) n_i + \frac{1}{2}\sum_{i\not=j}(n_i-K)(n_j-K)U_{ij}~.
\end{equation}
Here $n_i$ is the occupation number of site $i$ and $\mu$ 
is the chemical potential. The Coulomb interaction $U_{ij} = e^2/r_{ij}$
remains long-ranged since screening breaks down in the insulating phase. 
We set the interaction strength of nearest neighbor sites to 1 which fixes the
energy scale.  
The random potential values $\varphi_i$ are chosen 
independently from a box
distribution of width $2 W_0$ and zero mean.
The physics of the classical Coulomb glass model has been investigated 
by several analytical and numerical methods and its properties 
are comparatively well understood by now \cite{cgnum}
although the nature of the transport mechanism is still 
controversially discussed \cite{pollak92}. One of the remarkable 
features is the power-law gap in the zero-temperature
single-particle density of states (DOS) which is 
called the Coulomb gap \cite{es75,efros76}.

Since experiments deep in the insulating regime are difficult to carry out
most results on disordered insulators have been obtained from samples
not too far away from the metal-insulator transition \cite{exp}. Here the (single-particle) localization
length is still much larger than the typical distance between two sites
and the description of the electrons in terms of classical point charges becomes
questionable.
In order to investigate the influence of finite overlap between the states on
the properties of the insulating phase we have defined \cite{hf} the quantum 
Coulomb glass model, the minimal model of the "quantum insulating regime" 
which accounts for disorder, long-ranged interactions and the quantum nature of 
the electrons. It is obtained from the classical Coulomb glass 
by adding hopping matrix elements of strength $t$ between nearest neighbors.
The Hamiltonian of the quantum Coulomb glass reads
\begin{equation}
H =  -t  \sum_{\langle ij\rangle} (c_i^\dagger c_j + c_j^\dagger c_i) + H_{\rm cl},
\label{eq:Hamiltonian}
\end{equation}
where $c_i^\dagger$ and $c_i$ are the electron creation and annihilation operators
at site $i$, respectively,  and the sum runs over all pairs of nearest neighbor sites. 
In the limit $t \rightarrow 0$ the model (\ref{eq:Hamiltonian}) reduces to the classical
Coulomb glass, for vanishing Coulomb interaction but finite overlap it reduces to the
usual Anderson model of localization.

\section{Localization in an interacting system}
\label{sec:III}
It has been known for a long time \cite{andorg} that disorder leads to spatial localization
of single-particle states in a non-interacting system.  Since a system with
localized states at the Fermi level is insulating, the transition between
delocalized and localized states corresponds to a metal-insulator transition.
There is, however, much experimental and theoretical evidence that 
a description of disordered electronic systems in terms of non-interacting
particles is inadequate. For this reason one would
like to apply the concept of localization to interacting systems.
Unfortunately, a direct generalization of the non-interacting case is 
not possible since single-particle states are not defined in an 
interacting system while the many-particle states always correspond
to extended charge distributions. 
The deeper reason for these difficulties is, of course,
that in a non-interacting system the 
many-particle properties are completely determined by  the single-particle properties
whereas  the same is not true for interacting systems. 
Therefore, in a many-particle system one can consider several types of
"localization" (for single-particle or different many-particle excitations) 
which are all unrelated a priori.
In this section we discuss some of these ideas and we 
also present results for the corresponding physical quantities 
of a two-dimensional quantum Coulomb glass.

From an experimental point of view the most natural quantity 
to consider is probably the conductance since it is easily
measurable and its behavior determines whether the system 
is metallic or insulating. 
However, since the conductance is given by a 
two-particle Greens function and involves complicated 
zero-temperature and zero-frequency limits it is difficult to calculate
numerically \cite{phase}. 

In the following we concentrate on  {\em single-particle localization} 
which is the most direct generalization of Anderson localization to many-body systems.
Experimentally, single-particle localization should be reflected in 
the tunneling response of the system rather than in transport coefficients.

The simplest measure of Anderson localization for a single-particle state $|n\rangle$ is the
participation number $P$, defined as the inverse second moment of the spatial
probability distribution
\begin{equation} 
P^{-1}_n = \sum_j |\langle n | j \rangle |^4
\end{equation}
where the sum runs over all sites $j$.
In practice it is often averaged over all states with a certain energy $\varepsilon$
\begin{equation}
P^{-1}(\varepsilon) = \frac 1 {g(\varepsilon)}\: \frac 1 N \sum_n P^{-1}_n\: \delta(\varepsilon-\varepsilon_n)~.
\label{eq:partnum}
\end{equation}
A consistent generalization of this quantity to interacting systems
should fulfill at least the following conditions: (i) it should be well defined for 
any many-particle state and (ii) it should reduce to (\ref{eq:partnum})
for non-interacting electrons. Moreover, the desired quantity should
(iii) capture the physical idea of spatial localization and (iv) it should be 
easy to calculate. 

It has been suggested to define localization in a many-particle system via
the spatial distribution of the charge difference between the  
states of the same many-particle system with $N$ and $N+1$ particles, 
respectively. While this quantity fulfills the above
conditions (i), (ii), and (iv) it turns out that it is {\em not} a useful measure of localization in a 
disordered interacting system. The reason is that adding an extra electron to
a disordered interacting system will very often not only add some charge at a
few sites but completely rearrange the distribution of all electrons due to the
frustration introduced by the competition between disorder and interaction. Thus
simply calculating the participation number of the extra charge leads to an 
overestimation of delocalization. This can already be seen at the example of the 
classical Coulomb glass where we know that the electrons are completely
localized. Nevertheless the ground states of systems with $N$ and $N+1$
particles can be drastically different so that the charge difference between the two
is distributed on more than one site, effectively giving a participation number larger
than one. In general, the method will always fail, if adding an extra electron leads to
a {\em decrease} of the charge at some particular site since then the charge density
difference is not a proper probability distribution anymore (see Fig. \ref{fig:efros}).
\begin{figure}
\epsfxsize=\figuresize
\centerline{\epsffile{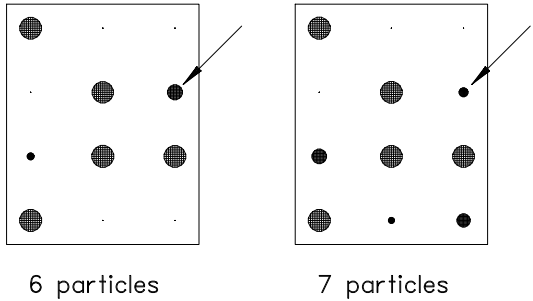}}
\caption{Comparison of the ground states of quantum Coulomb glasses
      with 6 and 7 particles. The size of the circles is proportional to the
      charge density at the site. Note that adding an extra electron {\em reduces}
       the charge density at the site marked by an arrow.}
\label{fig:efros}
\end{figure}

There is, however,  another quantity which fulfills conditions (i), (ii), and (iii)
for a generalization of the inverse participation number. In particular,
it nicely captures the physical idea of localization. This quantity is the
probability $R_p$ for an electron to return to its starting site in
infinite time. The energy-dependent return probability
 can be expressed in terms of single-particle Greens functions
\begin{equation}
R_p(\varepsilon) = \frac 1 N \sum_j \lim_{\delta \to 0} \frac \delta \pi \, G_{jj}(\varepsilon + i \delta)
     \, G_{jj}(\varepsilon - i \delta).
\end{equation}
For non-interacting electrons $P^{-1}(\varepsilon)=R_p(\varepsilon)$.
We note, however, that calculating $R_p$ requires the knowledge of 
all eigenstates of the Hamiltonian which makes this quantity
numerically expensive.
In Fig. \ref{fig:rp} we show the numerically determined return probabilities 
of the quantum Coulomb glass and the Anderson model.
\begin{figure}
\epsfxsize=\figuresize
\centerline{\epsffile{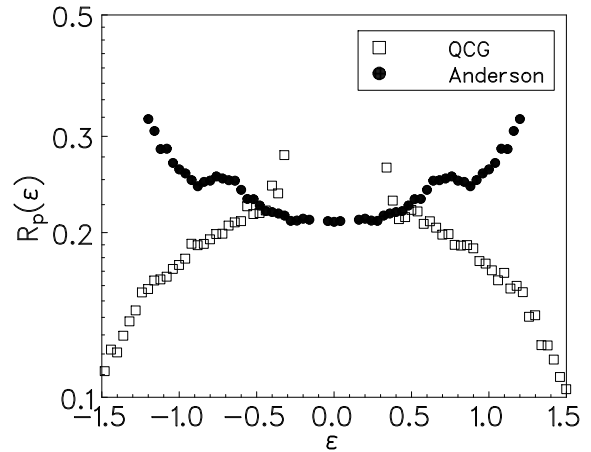}}
\caption{Return probability for a quasiparticle in the quantum Coulomb glass
 and the Anderson model on a 3x4 lattice, $t=0.3$, $W_0=1$, $K=0.5$}
\label{fig:rp}
\end{figure}
The two systems behave very differently. 
For the Anderson model we obtain the well-known behavior of
$P^{-1}(\varepsilon)$, viz. a minimum in the
band center and higher values in the tails. 
In the quantum Coulomb glass the
return probability has a maximum at the Fermi energy
and decreases quickly with increasing distance from the
Fermi energy. 
Thus we conclude that in the quantum Coulomb glass
single-particle excitations away from the Fermi energy tend to delocalize
while the excitations close to the Fermi energy which
dominate the low-temperature physics tend to localize
\cite{shklov}.
We have investigated the values of $R_p$ close to the Fermi level
for different values of the overlap $t$. In Fig. \ref{fig:rpt}
we show the resulting dependence and the corresponding
data for the Anderson model.
\begin{figure}
\epsfxsize=\figuresize
\centerline{\epsffile{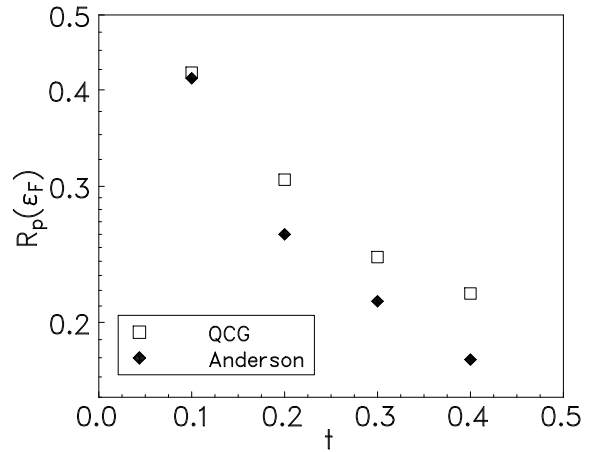}}
\caption{Return probability for a quasiparticle at the Fermi energy 
       as a function of $t$. Parameters are as in Fig. \protect\ref{fig:rp}.}
\label{fig:rpt}
\end{figure}
The figure shows that for  $t>0.1$ the interactions lead to
stronger localization in agreement with the results of our Hartree-Fock
calculation \cite{hf}. For $t=0.1$, quantum Coulomb glass and Anderson
model show identical behavior within the 
numerical accuracy. Therefore, we cannot exclude that
for very small $t$, 
the behavior may turn around, i.e., the interactions may favor delocalization, 
see  Ref. \cite{talamantes} and the discussion of Fock space localization below.

The return probability discussed above deals with the localization of a single
quasiparticle. As already mentioned, in an interacting system
many-particle properties cannot in general be derived from single-particle ones.
Therefore, to get a complete understanding of the localization 
properties one has to investigate many-particle quantities, too \cite{tip}.
One approach which is in some sense complementary to the 
study of the return probability is to analyze the localization properties
of the many-particle states with respect to a Fock-space basis set $\{|\alpha\rangle\}$.
Here the central quantity is the Fock-space participation number $P_F$
of a many-body state $|\nu\rangle$ which is given by
\begin{equation}
P_F^{-1}(\nu) = \sum_\alpha \langle \alpha | \nu\rangle^4.
\end{equation}
If we chose the Fock space basis to consist of Slater determinants
of site basis functions, $|\alpha\rangle = c_i^\dagger \ldots c_j^\dagger |0\rangle$
then $P_F=1$ for completely localized electrons and $P_F>1$ if the
electrons can move. A measure like this has already been used to 
characterize the influence of long-range interactions on Anderson localization
\cite{talamantes}.  
In Fig. \ref{fig:pfock} we present our results for the Fock-space
participation numbers of the quantum Coulomb glass and the Anderson 
model.
\begin{figure}
\epsfxsize=\figuresize
\centerline{\epsffile{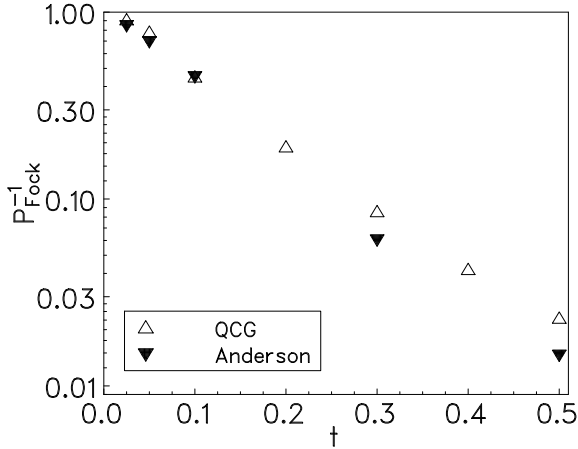}}
\caption{Fock space participation numbers for the ground states of the quantum
             Coulomb glass and the Anderson model. Parameters are as in Fig. \protect\ref{fig:rp}.}
\label{fig:pfock}
\end{figure}
The data show the same tendency as the return probabilities at the Fermi level
(Fig.\ref{fig:rpt}): For larger overlap $t$ the interactions lead to stronger localization
while for small $t$ we cannot find a statistically significant difference
between the quantum Coulomb glass and the Anderson model.
Thus, a change in the behavior at very small $t$ cannot be excluded.
This would resolve the seeming disagreement with Ref. \cite{talamantes}, where
Fock space localization in a related model was investigated for very
small overlaps, and the authors found a delocalizing influence of the interactions.
We note that, in principle, the concept of localization in Fock space can be
applied to any problem in which the Hamiltonian can be decomposed into
a reference part and a perturbation. Delocalization in Fock space then 
describes how the perturbation mixes the original eigenstates.
In a recent letter \cite{altshuler} this concept has been applied to 
explain the transition in the width of excited states measured in
tunneling conductance experiments in quantum dots \cite{sivan}.

Let us further mention that the localization transitions discussed above
also lead to transitions of the statistics of the corresponding energy
levels \cite{hf,talamantes,levshklov,levhof}.  
Analyzing the transitions of the level statistics is often 
numerically easier than dealing with the electron states itself.

\section{Conclusions}
In this paper we have discussed the generalization of Anderson localization to 
interacting systems. We first have considered the localization of single-particle 
excitations which can be described in terms of the return probability $R_p$.
Experimentally, this quantity should be reflected in the tunneling response of the 
system. We have also discussed the concept of Fock space localization
for disordered interacting electron systems.
As an example we have presented some numerical results for the quantum Coulomb
glass model of disordered insulators. 
In this concluding section we will discuss some aspects that have not yet been
covered. First, the quantum Coulomb glass is a model of spinless particles.
In one-dimensional systems there is, however, some evidence that the
electron spin plays an important role in determining the 
behavior of interacting disordered electrons \cite{giamarchi}. 
Therefore, in order to describe real electrons, including the spin into the
quantum Coulomb glass will be necessary in the future.
Second, our numerical examples were for very small lattice sizes.
In order to quantitatively analyze the behavior of disordered interacting 
electrons we have to extend the calculations to larger lattices. 
This can either be done by making approximations such as Hartree-Fock
\cite{hf} or by developing better numerical algorithms. Third, one of the main 
tasks of the future will be to establish relations, if any, between the different
types of localization and between the different quantities discussed here.

This work was supported in part by the DFG under grant
nos. Schr231/13-1, Vo659/1-1 and SFB393 and by the NSF under grant no. 
DMR-95-10185.



\begin{references}

\bibitem{reviews}For reviews see e.g. P. A. Lee and T. V. Ramakrishnan, Rev. Mod. Phys. {\bf 57}, 
       287 (1985), D. Belitz and T. R. Kirkpatrick, Rev. Mod. Phys. {\bf 66}, 261 (1994).
\bibitem{anderson}For a recent review see B. Kramer and A. MacKinnon, Rep. Progr. Phys. 
       {\bf 56}, 1469 (1993).
\bibitem{pollak70}M. Pollak, Discuss. Faraday Soc. {\bf 50}, 13 (1970).
\bibitem{es75}A. L. Efros and B. I. Shklovskii, J. Phys. C {\bf8}, L49 (1975).
\bibitem{efros76}A. L. Efros, J. Phys. C {\bf 9}, 2021 (1976). 
\bibitem{cgnum}Some recent results can be found in:  A. M\"obius, M. Richter and B. Drittler, Phys. 
          Rev. B {\bf 45}, 11568 (1992); T. Vojta, W. John and M. Schreiber,
          J. Phys. Condens. Matter {\bf 5}, 4989 (1993); M. Sarvestani, M. Schreiber and T. Vojta,
          Rev. B {\bf 52}, R3820 (1995).
\bibitem{pollak92}M. Pollak, Phil. Mag. B{\bf 65}, 657 (1992).
\bibitem{exp}See, e.g., K. M. Abkemeier, C. J. Adkins, R. Asal, and E. A. Davis, J. Phys. Condens. Matter {\bf 4}, 
          9113 (1992); A. Aharony, Y. Zhang and M. P. Sarachik, Phys. Rev. Lett. {\bf 68}, 3900 (1992);
          J. G. Massey and M. Lee, Phys. Rev. Lett. {\bf 75}, 4266 (1995).
\bibitem{hf}F. Epperlein, M. Schreiber and T. Vojta, Phys. Rev. B (1997), in press, cond-mat/9704068; 
         and contribution in this volume.
\bibitem{andorg}P. W. Anderson, Phys. Rev. {\bf 109}, 1492 (1958).
\bibitem{phase}There is a relation between the conductance and the sensitivity
       of the ground state energy to twisted boundary conditions, see \protect\cite{scal}, which
       can be used to calculate the conductance numerically. In dimensions larger
       than one, however, this method has its own problems.
\bibitem{scal}D. J. Scalapino, S. R. White, S. C.  Zhang, Phys. Rev. B {\bf 47}, 7995 (1993).
\bibitem{talamantes}J. Talamantes, M. Pollak, and L. Elam, Europhys. Lett. {\bf 35}, 511 (1996).
\bibitem{shklov}Note that the opposite was suggested by 
       I. L. Aleiner and B. I. Shklovskii, Int. J. Mod. Phys. B 8, 801 (1994).
\bibitem{tip}There has been a very active line of research considering the localization of just
           two interacting electrons. While this is not a many-particle problem, it tries to 
           capture essential aspects of the interplay of localization and interactions; see:
           D. L. Shepelyansky, Phys. Rev. Lett. {\bf 73}, 2607
          (1994); Y. Imry, Europhys. Lett. {\bf 30}, 405 (1995);
          D. Weinmann, A. M\"{u}ller-Groeling, J. L. Pichard, and
          K. Frahm, Phys. Rev. Lett. {\bf 75}, 1598 (1995);
          F. v. Oppen, T. Wettig, and J. M\"uller, Phys. Rev. Lett.  {\bf 76}, 491 (1996);
          R. A. R\"omer and M. Schreiber, Phys. Rev. Lett. {\bf 78}, 515 (1997).
\bibitem{altshuler} B. L. Altshuler, Y. Gefen, A. Kamanev, and L. S. Levitov,
         Phys. Rev. Lett {\bf 78}, 2803 (1997).
\bibitem{sivan}U. Sivan, F. P. Milliken, K. Milkove, S. Rishton, Y. Lee, J.M. Hong, V. Boegli,
          D. Kern, and M. DeFranza, Europhys. Lett. {\bf 25}, 605 (1994).
\bibitem{levshklov}B. I. Shklovskii, B. Shapiro, B. R. Sears, P. Lambrianides, and
          H. B. Shore,  Phys. Rev. {\bf 47}, 11487 (1993).
\bibitem{levhof}E. Hofstetter and M. Schreiber, Phys. Rev. B {\bf48}, 16979 (1993).
\bibitem{giamarchi} T. Giamarchi and H. J. Schulz, Phys. Rev. B {\bf 37}, 325 (1988).

\end{references}
\end{document}